\documentclass[runningheads]{llncs}
\usepackage[T1]{fontenc}
\usepackage{amsmath}
\usepackage{listings}
\usepackage{graphicx}
\usepackage[hidelinks]{hyperref}
\usepackage{color}

\usepackage{tikz}
\usetikzlibrary{arrows.meta,positioning, calc, shapes.geometric}

\title{WADE: An LSTM-Based Anomaly Detection Engine for WebAssembly Applications}
\titlerunning{WADE}

\author{Ilaria Campestrini$^{\star \dagger}$ \and Stefano Zanero$^{\star}$ \and Engin Kirda$^{\dagger}$}
\authorrunning{I. Campestrini et al.}

\institute{$^{\star}$ Politecnico di Milano, Milan, Italy \\
$^{\dagger}$ Northeastern University, Boston, MA, USA}

\begin{document}
\pagestyle{plain}

\maketitle
\thispagestyle{plain}

\begin{abstract}
WebAssembly (Wasm) enables near-native performance for web applications through a portable and sandboxed execution model. However, its limited runtime protection mechanisms (e.g., the lack of stack canaries and address space layout randomization (ASLR)) leave Wasm applications vulnerable to traditional attacks.

We present WADE (WebAssembly Anomaly Detection Engine), a system that detects anomalous runtime behaviors in browser-executed WebAssembly applications. WADE monitors function-level execution traces collected under benign conditions and trains a Long Short-Term Memory (LSTM) network to learn the normal execution behavior of each application. At runtime, the model flags deviations from these learned patterns as potential attacks.
Unlike signature-based approaches, WADE builds application-specific behavioral models, enabling detection of both known and previously unseen attacks without per-vulnerability signatures.

We evaluate WADE on a dataset of 67 WebAssembly applications spanning ten vulnerability classes, achieving zero false negatives and a 0.42\% false positive rate.

\end{abstract}

\section{Introduction}
WebAssembly (Wasm) has rapidly evolved into a key technology for high-perfor\-mance web applications. It provides a portable, low-level binary format that enables near-native execution speeds in the browser, supporting a wide range of use cases (e.g., from 3D rendering and gaming to cryptographic computations and machine learning inference) directly on the client side. Its efficiency, combined with language-agnostic compilation support, has made WebAssembly an integral component of modern web platforms.

Although Wasm was designed with security in mind, it still exposes a non-negligible attack surface. Its sandboxing model enforces isolation between the executing module and the host environment, and its type-safe interface prevents certain classes of memory violations. However, several critical runtime protection mechanisms commonly available in native environments are absent. As a result, traditional memory-corruption attacks, including buffer overflows and format-string vulnerabilities, can still subvert a WebAssembly module’s control flow and compromise its intended execution logic within the sandbox. 

As WebAssembly gains adoption in performance-critical and security-sensitive contexts (e.g., browser-based cryptographic libraries, blockchain clients, and enterprise web tools), the lack of advanced runtime protection mechanisms becomes an increasingly pressing concern. Detecting and mitigating malicious behavior within Wasm applications thus represents an important and timely research challenge.

We address this gap with WADE (WebAssembly Anomaly Detection Engine), a system for detecting anomalous execution behaviors in browser-based WebAssembly applications. WADE monitors function-level execution traces and trains an LSTM model on benign sequences to learn the normal behavior of each application.

Signature- and rule-based defenses require manually specifying allowed behaviors. For modern Wasm applications, the space of legitimate execution paths is large and input-dependent, making comprehensive rule construction brittle and difficult to maintain. We instead adopt a learning-based approach that models normal function-level execution sequences from benign traces and flags unlikely deviations, enabling detection of previously unseen exploits without per-vulnerability signatures.

\subsection{Threat model}
We focus on attacks against vulnerable WebAssembly applications where 
exploitation manifests at the function level, manipulating the Wasm call 
stack or redirecting indirect calls to alter the sequence of functions 
the runtime invokes.

The attacker interacts with the application through its legitimate input 
channels, such as user interfaces, file uploads, or network requests, but 
cannot alter the browser environment or the WebAssembly runtime. The 
attacker may possess full knowledge of the application, including its 
source code or binary, and can craft malicious inputs to trigger latent 
vulnerabilities in the compiled Wasm module.

Logic flaws or data-only attacks that do not affect execution patterns 
are out of scope, as they do not produce observable deviations in 
function-level traces.

\subsection{Contributions}
The main contributions of this paper are:

\begin{itemize}
    \item We propose WADE, an anomaly detection system for WebAssembly applications that identifies malicious control-flow behaviors from function-level traces collected in the browser.

    \item We design an LSTM model trained on function-level execution traces to learn per-application normal behavior.

    \item We develop a multi-threshold detection mechanism with four parallel detector instances and adaptive weighted voting, enabling robust identification of diverse attack manifestations without prior knowledge of attack patterns.

    \item We evaluate WADE on 67 WebAssembly applications spanning ten vulnerability classes, achieving zero false negatives and a 0.42\% false positive rate.

\end{itemize}

\section{Background}
\label{ch:background}

\subsection{WebAssembly}

WebAssembly (Wasm)~\cite{wasmsite,wasm} is a low-level, platform-independent binary instruction format introduced by the W3C in 2015 to enable portable, high-performance applications across heterogeneous execution environments. Rather than being written directly by developers, Wasm primarily serves as a compilation target for high-level languages such as C, C++, Rust, and Go. Toolchains like Emscripten~\cite{emscripten} translate LLVM intermediate representation into Wasm modules and provide a lightweight runtime layer that emulates system interfaces such as POSIX-compatible interfaces and Web APIs. This compilation model allows existing native codebases to be efficiently ported to the web or standalone Wasm runtimes while maintaining near-native performance.

A Wasm binary is organized into \emph{modules} defining functions, globals, tables, and linear memories, which can be imported from or exported to other modules. At runtime, each module is instantiated with its own mutable state - e.g., the operand stack, local variables, and linear memory - and executed on a stack-based virtual machine. Wasm functions consist of typed instructions operating on the operand stack. Direct function invocations use the \texttt{call} instruction, while indirect calls employ \texttt{call\_indirect}, which dynamically resolves a function index in a function table. Before execution, Wasm enforces static type and bounds checking to ensure that indirect calls match their declared signatures and that memory accesses remain within allocated regions. Any invalid access or type mismatch triggers a deterministic \emph{trap} that terminates execution.

WebAssembly provides no ambient access to the computing environment in which code is executed. Any interaction with the host system (e.g., I/O, file access, or operating system calls) must be explicitly performed through functions imported into the module by the host, also known as the \emph{embedder}. In the browser, Wasm modules are loaded, compiled, and invoked through a JavaScript API, whereas other environments such as Node.js or standalone Wasm runtimes expose their own host APIs for tasks like filesystem access or networking. To reduce this fragmentation and offer a unified programming model, the WebAssembly System Interface (WASI)~\cite{wasi} was proposed as a set of standard APIs providing secure, portable access to system resources.

Wasm programs operate on a single \emph{linear memory}, a contiguous array of bytes addressed by 32-bit offsets. This memory is untyped and unmanaged by the runtime: compilers partition it into regions for static data, heap, and stack-like areas managed at the program level. The virtual machine enforces only global memory bounds, leaving finer-grained memory integrity entirely under the control of the compiled code. This design choice, while efficient, means that WebAssembly applications written in unsafe languages such as C or C++ inherit the same memory vulnerabilities as their native counterparts.
 
Although Wasm enforces structured control flow and type safety, it lacks many runtime protections standard in native environments. Features such as stack canaries, Address Space Layout Randomization (ASLR), Data Execution Prevention (DEP), guard pages, unmapped pages, and memory protection flags are either absent or fundamentally incompatible with WebAssembly’s execution model. For example, native environments rely on virtual memory with \textit{unmapped pages} and \textit{guard pages} to crash a program upon out-of-bounds accesses, but WebAssembly’s linear memory is a single contiguous region without holes, meaning that any address within 
[0, max\_mem] is always readable and writable. This design prevents the insertion of guard pages between static data, the stack, and the heap, making buffer and stack overflows significantly more dangerous. Likewise, native virtual memory allows setting \textit{protection flags} (read-only, write-only, executable), whereas all data in WebAssembly linear memory is permanently writable, enabling corruption of critical program data. WebAssembly also does not support \textit{ASLR}, leaving stack and heap locations deterministic and predictable. Moreover, while \textit{DEP} is largely unnecessary in WebAssembly because linear memory is inherently non-executable, the absence of \textit{stack canaries} (Stack Smashing Protection, SSP) remains a significant omission, as WebAssembly intentionally excludes this mitigation despite its widespread use in modern native compilers. Finally, the type-checked Control-Flow Integrity (CFI) is coarse-grained, 
permitting redirection to any function that matches the expected type signature 
regardless of execution context. Consequently, while the sandbox prevents direct compromise of the host system, 
attacks such as heap corruption, indirect call redirection, and intra-module 
memory exploits can still occur within a WebAssembly instance, allowing 
attackers to exfiltrate sensitive data, bypass security checks, manipulate 
application logic, or inject malicious content into the browser DOM. These limitations motivate ongoing research on control-flow integrity, memory safety instrumentation, and runtime anomaly detection for WebAssembly applications.

\subsection{Related Work}

Early studies demonstrated that WebAssembly’s sandboxing does not eliminate traditional binary exploitation.

Lehmann et al.~\cite{lehmann} showed that memory corruption vulnerabilities such as buffer overflows and use-after-free remain exploitable within Wasm modules, allowing manipulation of linear memory and function tables despite structured control flow.

Massidda et al.~\cite{vuln_wasm} were the first to demonstrate that format string vulnerabilities can be exploited in WebAssembly, enabling memory disclosure and arbitrary writes through the linear memory model. They also presented practical exploits achieving arbitrary code execution within the Wasm sandbox, showing that classic exploitation chains remain viable despite the isolation model.

Stiévenart et al.~\cite{security_risks} extended this analysis by studying unsafe C applications compiled to WebAssembly, finding that memory-safety issues are directly preserved during compilation. Their results confirm that WebAssembly does not inherently mitigate vulnerabilities present in the source language.

Perrone and Romano~\cite{perrone} provided a comprehensive review of WebAssembly security research and identified a strong imbalance: among thirteen detection systems, most target cryptojacking attacks, while none address memory corruption or control-flow hijacking.

Heinrich et al.~\cite{heinrich} proposed one of the first anomaly-detection approaches for WebAssembly, modeling runtime traces as categorical data and classifying them using traditional machine learning models such as decision trees and random forests. However, their method relies on a single global model across applications and simple, input-free benchmarks, limiting scalability and accuracy in complex browser environments.

A recent work by Vogel et al.~\cite{wasmeye} presents WasmEye, a platform-independent anomaly detection system that embeds ensemble learning 
models directly into WebAssembly modules. WasmEye uses bag-of-words 
representation, counting function calls and memory accesses within fixed time 
windows, as input to an ensemble of four machine learning models (One-Class SVM, 
Isolation Forest, Local Outlier Factor, and Elliptic Envelope).
Their evaluation focuses on IoT/edge scenarios with three synthetic benchmarks (arithmetic operations, a SAT solver, and SQLite)
plus one real-world vulnerable program. Our work significantly differs in four key aspects: (1) we target 
browser-based WebAssembly specifically; (2) our LSTM-based approach models function call sequences and their dependencies, 
detecting anomalous orderings that frequency-based counting cannot identify; (3) our LSTM processes execution 
incrementally at each function call rather than in batches, enabling online detection as execution unfolds; and (4) we evaluate on 
67 vulnerable programs across 10 vulnerability classes, demonstrating broader 
applicability.

In summary, existing research has thoroughly analyzed WebAssembly vulnerabilities, but offers limited detection mechanisms. Most detection efforts focus on narrow attack classes or simplified application settings, leaving open the challenge of identifying memory-corruption and control-flow attacks in real-world browser-based WebAssembly programs.

\section{Design}
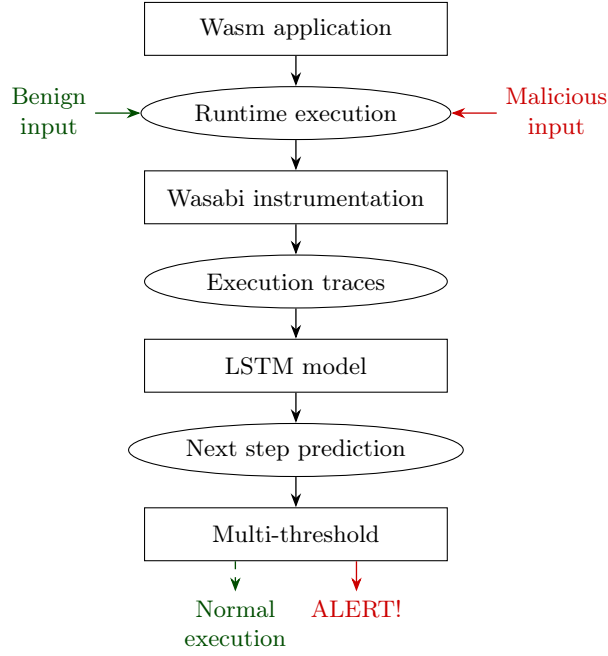
\begin{figure}[!t]
\centering
\begin{tikzpicture}[
    node distance=0.4cm,
    every node/.style={font=\small},
    block/.style={
        rectangle,
        draw,
        minimum width=4cm,
        minimum height=0.7cm,
        align=center,
        text=black
    },
    oval/.style={
        ellipse,
        draw,
        minimum width=4cm,
        minimum height=0.7cm,
        align=center,
        text=black
    },
    arrow/.style={->, >=Stealth, line width=0.5pt, black}
]
\node[block] (wasm) {Wasm application};
\node[oval,  below=of wasm]   (runtime) {Runtime execution};
\node[block, below=of runtime] (wasabi) {Wasabi instrumentation};
\node[oval,  below=of wasabi]  (traces) {Execution traces};
\node[block, below=of traces]  (lstm) {LSTM model};
\node[oval,  below=of lstm]    (next) {Next step prediction};
\node[block, below=of next]    (detect) {Multi-threshold};

\node[left=0.6cm of runtime, align=center, text=green!30!black] (benign) {Benign\\input};
\node[right=0.6cm of runtime, align=center, text=red!80!black] (malicious) {Malicious\\input};

\node[below=0.4cm of detect, xshift=-0.8cm, text=green!30!black, align=center] (normal) {Normal\\execution};
\node[below=0.4cm of detect, xshift= 0.8cm, text=red!80!black, align=center]  (alert)  {ALERT!};

\draw[arrow] (wasm)   -- (runtime);
\draw[arrow] (runtime)-- (wasabi);
\draw[arrow] (wasabi) -- (traces);
\draw[arrow] (traces) -- (lstm);
\draw[arrow] (lstm)   -- (next);
\draw[arrow] (next)   -- (detect);
\draw[arrow, green!30!black] (benign) -- (runtime);
\draw[arrow, red!80!black] (malicious) -- (runtime);

\draw[arrow, green!30!black, dashed] 
    ($(detect.south) + (-0.8cm,0)$) -- (normal);
\draw[arrow, red!80!black]   
    ($(detect.south) + (0.8cm,0)$) -- (alert);
\end{tikzpicture}
\caption{WADE components overview.}
\label{fig:overview}
\end{figure}

WADE operates in two phases: offline training and online detection. During training, benign execution traces are collected from the target application and used to train an LSTM model to learn normal execution patterns. During detection, the trained model monitors live browser execution and flags deviations in real time.
Figure~\ref{fig:overview} provides a high-level overview of the WADE architecture during detection. Each WebAssembly application is instrumented with Wasabi, which monitors its execution in real time. As the application runs, Wasabi records the sequence of function invocations, producing a function-level execution trace incrementally.
WADE feeds this stream to the LSTM one function call at a time. At each step, the model predicts the next function and computes the corresponding prediction error. If the error exceeds the predefined thresholds, WADE raises an alert; otherwise, execution continues uninterrupted.

\subsection{Runtime Instrumentation}
Control-flow attacks in WebAssembly manifest through alterations in function 
execution patterns. Simply tracking which functions are called is insufficient:
attacks can reorder or skip legitimate functions. For instance, a validation 
bypass might redirect from \texttt{authenticate} to \texttt{process\_data}, 
omitting \texttt{validate}. Each function is legitimate, but the transition 
is anomalous. 

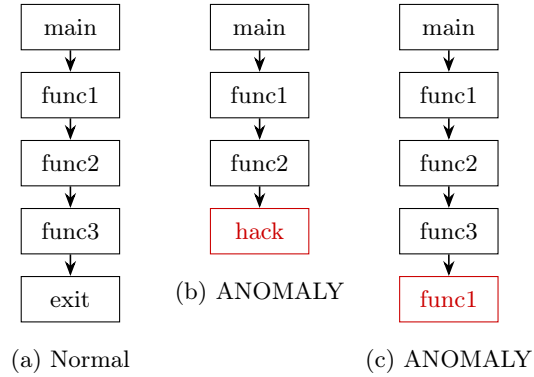
\begin{figure}[t]
\centering
\begin{tikzpicture}[
    node distance=0.9cm,
    every node/.style={draw, rectangle, minimum width=1.3cm, minimum height=0.6cm, font=\small},
    arrow/.style={->, >=Stealth, line width=0.6pt, black}
]

\node (main1) at (0,0) {main};
\node (f1a)   [below of=main1] {func1};
\node (f2a)   [below of=f1a]   {func2};
\node (f3a)   [below of=f2a]   {func3};
\node (exit1) [below of=f3a]   {exit};

\draw[arrow] (main1) -- (f1a);
\draw[arrow] (f1a) -- (f2a);
\draw[arrow] (f2a) -- (f3a);
\draw[arrow] (f3a) -- (exit1);

\node[below=0.2cm of exit1, draw=none, font=\small] {(a) Normal};

\node (main2) at (2.5,0) {main};
\node (f1b)   [below of=main2] {func1};
\node (f2b)   [below of=f1b]   {func2};
\node (hack)  [below of=f2b, draw=red!80!black, text=red!80!black]   {hack};

\draw[arrow] (main2) -- (f1b);
\draw[arrow] (f1b) -- (f2b);
\draw[arrow] (f2b) -- (hack);

\node[below=0.2cm of hack, draw=none, font=\small] {(b) ANOMALY};

\node (main3) at (5,0) {main};
\node (f1c)   [below of=main3] {func1};
\node (f2c)   [below of=f1c]   {func2};
\node (f3c)   [below of=f2c]   {func3};
\node (f1c2) [below of=f3c, draw=red!80!black, text=red!80!black]   {func1};

\draw[arrow] (main3) -- (f1c);
\draw[arrow] (f1c) -- (f2c);
\draw[arrow] (f2c) -- (f3c);
\draw[arrow] (f3c) -- (f1c2);

\node[below=0.2cm of f1c2, draw=none, font=\small] {(c) ANOMALY};

\end{tikzpicture}

\caption{Comparison of normal and anomalous call graphs.}
\label{fig:control-flow}
\end{figure}

Figure~\ref{fig:control-flow} illustrates this distinction with a different 
example. Normal execution 
follows expected control-flow patterns (a). Attacks may manifest by introducing 
calls to unexpected functions (b) or by reordering legitimate functions (c). 
To detect both types of anomalies, we must monitor not only which functions 
are called, but their \textbf{sequential ordering} and the transitions between them.

We extract function-level execution traces at runtime using \textbf{Wasabi}~\cite{wasabi}, a dynamic instrumentation framework for WebAssembly. Wasabi operates directly on Wasm bytecode, parsing the binary format and inserting lightweight hooks that invoke JavaScript callbacks before and after selected instructions.
This enables fine-grained runtime monitoring at the WebAssembly function level without 
requiring source code access, while preserving the original semantics by 
maintaining consistent stack and memory states while instrumented code executes. 

Wasabi supports several analyses, including instruction profiling, call-graph generation, and memory-access tracing. Such comprehensive instrumentation provides full visibility into program behavior at the cost of moderate overhead.  

For our system, we employ and extend Wasabi’s \textit{call-graph.js} analysis to record function-level traces. The script inserts hooks on function entry and exit, logging the sequence of calls. By operating at the function granularity rather than per-instruction, this approach substantially reduces runtime overhead while retaining the information required for control-flow–based anomaly detection. We further modify the script to support multi-input execution sessions, ensuring that each input produces an independent trace suitable for analysis across different runs.

For example, consider the following simple C program, \texttt{hello.c}:

\begin{footnotesize}
\begin{lstlisting}[language=C, basicstyle=\ttfamily, frame=none, breaklines=true]
void greet() {
    printf("Hello, World!\n");
}
int main() {
    greet();
    return 0;
}
\end{lstlisting}
\end{footnotesize}

The source code is compiled to WebAssembly using Emscripten ({\small \texttt{emcc hello.c -o hello.wasm}}), producing the module \textit{hello.wasm}. Wasabi instrumentation is then applied to this module, generating the following call graph:

\begin{footnotesize}
\begin{verbatim}
__wasm_call_ctors --> emscripten_stack_init 
__wasm_call_ctors --> 43 
43 --> 41 
41 --> 40 
main --> 4 
4 --> greet 
greet --> 7 
7 --> 35 
35 --> 26 
26 --> 11 
26 --> 27 
26 --> 18 
27 --> 28 
28 --> 25 
25 --> 8 
8 --> wasi_snapshot_preview1.fd_write 
8 --> 39 
25 --> 24 
__stdio_exit --> 15 
15 --> 14 
__stdio_exit --> 17
\end{verbatim}
\end{footnotesize}

This trace illustrates WebAssembly's runtime model: execution 
begins with initialization (\texttt{\_\_wasm\_call\_ctors}), 
proceeds through internal functions to \texttt{main} and 
\texttt{greet}. The expansion of \texttt{printf} reveals 
multiple abstraction layers before the actual I/O operation 
via the WASI system call \texttt{wasi\_snapshot\_preview1.\allowbreak fd\_write}, 
followed by cleanup (\texttt{\_\_stdio\_exit}).

\subsection{Trace representation}
\label{trace_representation}

We adopt a \textbf{sequential representation}: each execution trace is encoded as an ordered sequence of function identifiers matching the runtime call order. This preserves temporal relationships between calls, ensuring that two executions involving the same functions in different orders are treated as distinct behaviors. The representation sacrifices structural information such as call hierarchies, but retains the temporal dynamics essential for detecting execution deviations.

We represent each execution trace as a sequence by appending, for each 
edge $A \rightarrow B$, both $A$ and $B$ in order, explicitly preserving 
the source-destination pairs of each function call. 
Applying this encoding to the \textit{hello} example yields:
{\small
\texttt{
[\_\_wasm\_call\_ctors, emscripten\_stack\_init, \_\_wasm\_call\_ctors, 43, 43, 41, \\41, 40, main, 4, 4, greet, greet, 7, 7, 35, 35, 26, 26, 11, 26, 27, 26, \\18, 27, 28, 28, 25, 25, 8, 8, wasi\_snapshot\_preview1.fd\_write, 8, 39, 25, \\24, \_\_stdio\_exit, 15, 15, 14, \_\_stdio\_exit, 17]}}.

A more compact encoding could avoid repeating elements, but would introduce ambiguity. Maintaining explicit source-destination pairs ensures each function transition is unambiguously defined; the redundancy is deliberate and prioritizes correctness over space efficiency.

\subsection{Model}
Valid WebAssembly execution sequences exhibit complex temporal dependencies that are impractical to encode as manual rules. Rule-based approaches also become obsolete as applications evolve, requiring costly rewriting. We instead adopt a machine learning approach that automatically learns the sequential structure of normal execution from benign training data and requires only automated retraining when an application is updated.

Specifically, we employ a \textbf{Long Short-Term Memory} (LSTM) network~\cite{lstm}, which is well suited for this task because it processes sequences incrementally and maintains an internal state that evolves over time, capturing both short- and long-range dependencies between function calls. This is crucial for detecting anomalous control-flow patterns that span many calls and cannot be identified by models limited to local context, as we empirically confirm in Section~\ref{ch:evaluation}.
Unlike attention-based architectures that require the entire sequence to be available before processing, LSTMs naturally operate in streaming contexts where function calls are observed in real time. Their recurrent structure enables efficient sequential inference with low computational overhead, making them well suited for runtime anomaly detection directly within the browser.

\subsubsection{Input preprocessing}
\label{input_preprocessing}
Before training an LSTM model for a given program, we build a vocabulary of all functions in its WebAssembly module using \texttt{wasm-objdump} from the WebAssembly Binary Toolkit (WABT)~\cite{wabt}. Running \texttt{wasm-objdump -x hello.wasm} produces a textual description of the binary’s contents, including imports, functions, tables, memory, globals, exports, elements, and code sections.

Of particular interest are the \texttt{Import} and \texttt{Function} sections: the former lists functions the module expects from the host environment, the latter describes all functions defined inside the module.

We extract all available function identifiers, using symbolic names where 
available and falling back to numeric indices otherwise. Each function is 
assigned a vocabulary index, and a special end-of-sequence token 
(\texttt{<EOS>}) is appended. The resulting mapping constitutes the 
vocabulary. For example, from \textit{hello.wasm} we obtain a vocabulary 
of 59 functions. Applying this mapping to the symbolic sequence yields the 
following numerical sequence:
{\small \texttt{[44, 49, 44, 32, 32, 30, 30, 29, 53, 28, 28, 52, 52, 39, 
39, 23, 23, 14, 14, 1, 14, 15, 14, 7, 15, 16, \\16, 13, 13, 40, 40, 57, 40, 
27, 13, 12, 43, 5, 5, 4, 43, 6, 58]}}.

\subsubsection{Model architecture}
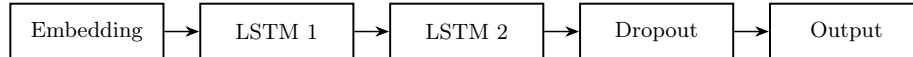
\begin{figure}[b]
\centering
\resizebox{\columnwidth}{!}{%
\begin{tikzpicture}[
    layer/.style={rectangle, draw=black, line width=0.8pt, minimum width=2.2cm, minimum height=0.8cm},
    arrow/.style={->, >=Stealth, line width=0.6pt, black}
]
\def\xdist{0.5}
\node[layer] (embedding) at (0,0) {Embedding};
\node[layer, right=\xdist of embedding] (lstm1) {LSTM 1};
\node[layer, right=\xdist of lstm1] (lstm2) {LSTM 2};
\node[layer, right=\xdist of lstm2] (dropout) {Dropout};
\node[layer, right=\xdist of dropout] (output) {Output};
\draw[arrow] (embedding.east) -- (lstm1.west);
\draw[arrow] (lstm1.east) -- (lstm2.west);
\draw[arrow] (lstm2.east) -- (dropout.west);
\draw[arrow] (dropout.east) -- (output.west);
\end{tikzpicture}
}
\caption{Model architecture.}
\label{fig:lstm_architecture}
\end{figure}

The architecture comprises four sequential components: an embedding layer, a two-layer stacked LSTM, a dropout layer, and a dense output layer. Figure~\ref{fig:lstm_architecture} illustrates the model architecture for \textit{hello.wasm}.
 
The model processes variable-length input sequences and produces a probability distribution over the vocabulary at each time step. It follows an autoregressive language modeling approach: each function call is treated as a token, and the model learns to predict the next function given all preceding calls.

The \textit{embedding layer} maps each function identifier to a dense representation, enabling the network to capture semantic relations between functions while reducing sparsity compared to one-hot encodings. The embedded sequence is then processed by two \textit{LSTM layers}, which model temporal dependencies and learn sequence relationships among tokens. The first LSTM layer (\textit{LSTM\textsubscript{1}}) processes the embedded input sequence and learns low-level sequential patterns, producing a hidden state sequence passed to the second layer. The second LSTM layer (\textit{LSTM\textsubscript{2}}) operates on this intermediate representation, capturing higher-level temporal dependencies and longer-range context between function calls. This stacking enables the model to learn hierarchical temporal abstractions of the execution trace, and was selected over a single-layer configuration based on preliminary experiments, which showed reduced sensitivity to long-range dependencies in the latter. A \textit{dropout layer} \cite{dropout} is applied after the recurrent layers to improve generalization and mitigate overfitting. Finally, a fully connected \textit{output layer} projects the last hidden state to the vocabulary dimension, producing a categorical probability distribution over possible next functions via a \textit{softmax} \cite{pytorch_softmax} activation.

The model was configured with an embedding dimension of 
32, a hidden size of 256 units per LSTM layer, and a dropout 
rate of 0.3. Training used the Adam optimizer with a learning 
rate of $10^{-3}$.

\subsubsection{Training}

The model is trained offline, exclusively on benign execution traces, to capture each application's normal control-flow behavior. Training makes a single pass over the collected traces: each trace is processed sequentially, with LSTM hidden states initialized to zero at the start of every sequence to prevent information leakage across traces.

At each time step $t$, the embedding corresponding to function $f_t$ is fed through the LSTM layers, producing a hidden state $h_t$ used to predict the probability distribution over the next function $f_{t+1}$. The model is optimized to minimize the \textit{categorical cross-entropy} loss~\cite{crossentropy}:

\begin{equation}
\mathcal{L} = - \sum_{t=1}^{S-1} \log P_\theta(f_{t+1} \mid f_1, \dots, f_t)
\end{equation}

where $P$ denotes the model's predicted probability distribution, $f_t$ is the function call at time step $t$, and $S$ is the sequence length. 

Gradients are accumulated throughout the entire sequence, and model parameters are updated only after the full trace has been processed, ensuring that learning exploits the complete temporal context of each sequence rather than isolated transitions.

\subsection{Anomaly Detection Mechanism}

The trained LSTM model is deployed for real-time anomaly detection. 
At each time step $t$, given the observed function sequence $f_{1:t}$, the model predicts the probability distribution over the next function call $\hat{p}_{t+1} = P_\theta(f_{t+1} \mid f_{1:t})$. 
The \textit{prediction error}, defined as the negative log-likelihood of the actual function, quantifies how unexpected the observation is. 
Benign executions produce low errors consistent with learned control-flow patterns, while attacks cause higher deviations.

To translate prediction errors into anomaly decisions, we design a multi-threshold detection mechanism that distinguishes between different degrees of deviation.  
Three thresholds are defined from the empirical distribution of prediction errors observed on benign training traces: 
\begin{enumerate}
    \item a \textit{spike threshold} ($\tau_{\text{spike}}$, 99.99th percentile) for extreme single-step anomalies
    \item a \textit{high threshold} ($\tau_{\text{high}}$, 99.95th percentile) for strong deviations
    \item a \textit{medium threshold} ($\tau_{\text{medium}}$, 99.9th percentile) for moderate persistent ano\-malies
\end{enumerate} 

Each threshold corresponds to a distinct type of anomaly and dictates how many consecutive violations are required to raise an alert.  
Errors above $\tau_{\text{spike}}$ immediately trigger an alarm after a single occurrence, as such events are virtually absent in benign executions.  
When errors fall between $\tau_{\text{high}}$ and $\tau_{\text{spike}}$, two consecutive violations are required, whereas medium-level deviations between $\tau_{\text{medium}}$ and $\tau_{\text{high}}$ must persist for at least three steps before being classified as anomalous.  
This design ensures robustness against isolated fluctuations while remaining sensitive to sustained deviations.

A particularly relevant case occurs when the execution encounters a function that was never observed during training.  
Since the model assigns near-zero probability to such unseen functions, the resulting error $\varepsilon_t$ sharply exceeds the spike threshold, leading to an immediate alert.  
This behavior is desirable, as the invocation of an unknown function often signals a control-flow hijacking or code injection attempt.   

Conversely, attacks may occur without introducing new functions, by reordering legitimate ones or skipping expected calls.  
In these cases, the model still recognizes each function individually, but detects inconsistencies in their ordering or temporal dependencies.  
Such violations typically generate medium or high errors, which require multiple consecutive anomalies before raising an alert.  

To improve robustness across diverse attack manifestations, WADE employs 
four parallel detector instances, each configured with a distinct sensitivity 
factor ($\alpha \in \{0.70, 0.90, 1.10, 1.30\}$) applied to scale the 
thresholds. Each instance independently evaluates the prediction error at 
every time step and casts a weighted vote. A combined weighted vote exceeding 
0.5 triggers an alert. The instance weights are updated dynamically at each step using 
exponential weighting: an instance is penalized if it votes anomalous 
when the error is below the recent mean (likely false positive), or 
votes normal when the error substantially exceeds it (likely false 
negative). The learning rate decays proportionally to $1/\sqrt{n}$, 
where $n$ is the number of steps processed. Over time, instances that 
make more errors receive lower weights, allowing the ensemble to adapt 
to the specific behavioral characteristics of each application.

This design enables the detector to tolerate benign fluctuations while remaining sensitive to meaningful deviations in control flow.

Since our model is trained exclusively on benign traces for each specific program, it cannot anticipate which attack patterns might emerge; it only learns what constitutes normal behavior for that particular application.  
This limitation necessitates an adaptive detection mechanism capable of identifying diverse manifestations of malicious activity without prior knowledge of their characteristics. The proposed multi-threshold strategy addresses this need by combining immediate responsiveness to critical anomalies with robustness against noise, effectively covering a wide spectrum of possible deviations observed in WebAssembly execution.

\section{Evaluation}
\label{ch:evaluation}

A key challenge in evaluating our system lies in obtaining a suitable dataset of vulnerable WebAssembly applications. The dataset must include programs that accept external input, enabling controlled triggering of vulnerabilities to generate both benign and malicious execution traces.

\subsection{Dataset}
Existing datasets such as \textit{Wasmbench}~\cite{wasmbench} and \textit{Wasmizer}~\cite{wasmizer} provide large collections of real-world WebAssembly binaries (8,461 and 8,915 respectively). However, these datasets lack explicit labeling of vulnerable and benign behaviors. In contrast, the \textit{Juliet Test Suite}~\cite{juliet} is the de facto benchmark for C/C++ vulnerability analysis, comprising 54,484 test cases across 118 CWEs. Stiévenart et al.~\cite{sac22} derived a WebAssembly version of Juliet containing 4,911 C programs compiled to x86 and WebAssembly across 39 CWE (Common Weakness Enumeration) categories. However, these test cases are typically minimal, and do not require external input, making them unsuitable for our evaluation.

Since no standardized benchmark for vulnerable WebAssembly binaries currently exists, we constructed a dedicated dataset for our evaluation. We started from 7 real-world vulnerable seed programs documented in established security research from Lehmann et al.~\cite{wasm_binary_security}, Massidda et al.~\cite{github_vulnwasm}, and Stiévenart et al.~\cite{sac22}.
To increase evaluation coverage and enable systematic testing across diverse code patterns, we used Claude Opus 4~\cite{claude} (Anthropic) to generate variant programs from each seed. 
For each seed program, we provided the LLM with the complete source code 
and the following instructions to generate variants:

\begin{scriptsize}
\begin{verbatim}
Generate a variant of the following vulnerable C program that satisfies these requirements:
1. Preserve the exact vulnerabilities present in the original code
2. Modify the functions without vulnerabilities
3. Substantially increase program complexity by adding additional 
   functions
4. Diversify control flow paths through varied conditional branches 
   and function call sequences
5. Maintain compilability to WebAssembly via Emscripten
Do not introduce any security mitigations. The vulnerable code 
pattern must remain unchanged.
[Complete seed program source code]
\end{verbatim}
\end{scriptsize}
For 6 of the 7 seed programs, we generated 10 variants each, yielding 
60 LLM-generated programs which, combined with the 7 seed programs, 
constitute our final dataset of 67 programs.

Every LLM-generated variant underwent rigorous manual validation before 
inclusion in the final dataset, following this process:
\begin{enumerate}
    \item We manually inspected the source code to verify that the vulnerable function remains unchanged.

    \item We compiled each variant to WebAssembly using Emscripten to verify syntactic correctness and successful binary generation.
    
    \item For each variant, we used targeted exploit inputs and verified successful exploitation.
\end{enumerate}

Variants that failed any validation criterion were rejected and regenerated. Analysis of rejected variants revealed mainly inadvertent security fixes (LLM replaced \texttt{strcpy} with \texttt{strncpy}, added bounds checking, or introduced input validation) and compilation failures.

This validation process ensured that our final dataset contains only programs with verified, exploitable vulnerabilities suitable for rigorous evaluation.

Our final dataset comprises 67 programs derived from seven seed applications, each representing one or more vulnerability classes. Table~\ref{table:dataset_composition} summarizes the coverage of vulnerability classes across our dataset. 

\begin{table} [b]
\caption{Vulnerability classes present in our dataset}
\label{table:dataset_composition}
\centering
\begin{tabular}{|l|c|}
\hline
\textbf{Vulnerability class} & \textbf{Programs in dataset} \\ \hline
Buffer Overflows & 34 \\ \hline
Stack-Based Buffer Overflows & 23 \\ \hline
Heap Metadata Corruption & 22 \\ \hline
Overwriting Heap Data & 22 \\ \hline
Overwriting Stack Data & 22 \\ \hline
Redirect Indirect Calls & 22 \\ \hline
Format String & 11 \\ \hline
Integer Overflow/Underflow & 11 \\ \hline
Cross-Site Scripting (XSS) & 1 \\ \hline
Code Injection & 1 \\ \hline
\end{tabular}

\smallskip
{\footnotesize Programs may span multiple vulnerability classes.}
\end{table}

In total, we managed to cover 10 out of 16 classes identified by Heinrich~\cite{phdthesis}. The six excluded 
categories are: stack overflow, overwriting constant 
data, server-side remote code execution, 
side-channel attacks, compiler bugs, and constant-time 
violations. These are excluded because they 
either do not produce observable control-flow anomalies (side-channels, 
constant-time violations), require 
alternative detection approaches (stack 
overflow, overwriting constant data), fall outside our browser-based deployment context 
(server-side remote code execution), or involve bugs introduced 
during C-to-WebAssembly compilation (compiler bugs).
This coverage focuses WADE on attacks that manifest as deviations in function execution 
patterns, consistent with our threat model.

Afterwards, we generated the corresponding training and test sets using two automated scripts that randomly produced input values to exercise as many execution paths as possible. 

The training set consists exclusively of benign traces, obtained by executing the instrumented WebAssembly modules with non-exploiting inputs. To ensure comprehensive coverage, we intentionally triggered benign error conditions during training data generation to capture error handling paths. 
The number of samples in the training set varies notably based on the complexity and type of program, ranging from 600 to 2000 samples.

The test set includes both benign and malicious traces, generated from exploiting and non-exploiting inputs. Although real-world attacks are rare, we intentionally balanced the dataset by increasing the proportion of malicious traces (7–10\%) to ensure stable and meaningful evaluation metrics~\cite{class_imbalance}. The size of the test set is approximately 15–20\% of the training set.

To support reproducibility, the dataset will be made publicly 
available upon publication.

\textbf{Realism of evaluation.} All traces (i.e., benign and malicious) were collected from real executions of the compiled WebAssembly binaries in a standard browser engine, rather than from abstract models or synthetic control-flow graphs. Malicious traces were generated by running working exploits against the same instrumented programs, ensuring that the observed anomalies reflect genuine exploitation effects at runtime. While we oversample malicious traces to stabilize evaluation metrics, this balancing changes only the class proportions and does not alter the underlying execution behaviors captured in the traces.

\subsection{Experimental Setup}
Trace collection was performed on a 
MacBook Pro (2019) with an Intel Core i7-8557U CPU, 8 GB RAM, running macOS 
Sequoia 15.5. We used Emscripten 3.1.6 for compiling C programs to WebAssembly, 
WABT 1.0.36 (wasm-objdump) for binary analysis, and Wasabi 0.3.0 for dynamic 
instrumentation. Trace collection was automated using Node.js 22.16.0 with 
Puppeteer 24 controlling a headless Chrome browser.

All machine learning experiments were conducted on a cloud-based GPU instance 
equipped with an NVIDIA Tesla T4 GPU (15.36 GB VRAM), an Intel Xeon CPU 
(4 cores @ 2.00 GHz), and 32 GB RAM, running Ubuntu 22.04.4 LTS. The software 
stack included Python 3.11, PyTorch 2.6 with CUDA 12.4 support, NumPy 1.26, 
and Pandas 2.2.

\subsection{Results}
The model performance was evaluated using standard metrics, including accuracy, precision, recall, specificity and F1-score \cite{powers}.
The average performance metrics are summarized in Table~\ref{tab:metrics}.

\begin{table} [t]
\caption{Average performance metrics on our dataset}
\label{tab:metrics}
\centering
\begin{tabular}{|l|c|}
\hline
\textbf{Metric} & \textbf{Value} \\ \hline
Accuracy & $99.62\%$ \\ \hline
Precision & $96.90\%$ \\ \hline
Recall & $100\%$ \\ \hline
Specificity & $99.58\%$ \\ \hline
F1-Score & $98.42\%$ \\ \hline
False Positive Rate & $0.42\%$\\ \hline
False Negative Rate & $0\%$\\ \hline
\end{tabular}
\end{table}

WADE achieved 100\% recall, detecting every attack across all 67 programs with zero false negatives. In anomaly detection this metric is critical, as missed attacks can have serious consequences.

Precision of 96.90\% confirms that false alarms remain rare, and the F1-score of 98.42\% reflects consistent performance across both metrics. These results hold across programs of varying complexity and vulnerability class.

Analysis of false positives revealed that the training scripts had missed certain benign execution paths, confirming that WADE behaves as a novelty detector: it flags unseen behaviors regardless of whether they are malicious. Insufficient training coverage therefore causes legitimate but unobserved paths to be misclassified, underscoring the importance of exercising all reachable code paths during training.

To empirically validate the choice of LSTM over simpler sequential 
models, we evaluated a Markov chain baseline on the same dataset. 
The Markov chain models function-call transitions as a first-order 
probability distribution, capturing only local pairwise dependencies 
between consecutive calls. Table~\ref{tab:baseline} reports the comparative results. The Markov chain achieves a recall of only 7.1\%, 
confirming that local-dependency models are insufficient for detecting 
control-flow attacks that span longer execution contexts.

\begin{table}[b]
\caption{Comparison between WADE and Markov chain baseline}
\label{tab:baseline}
\centering
\begin{tabular}{|l|c|c|c|c|c|}
\hline
\textbf{Method} & \textbf{Accuracy} & \textbf{Recall} & 
\textbf{Precision} & \textbf{Specificity} & \textbf{F1-Score} \\ \hline
Markov Chain & 82.1\% & 7.1\% & 50.0\% & 82.9\% & 12.5\% \\ \hline
WADE   & 99.62\% & 100\% & 96.90\% & 99.58\% & 98.42\% \\ \hline
\end{tabular}
\end{table}

Direct comparison with prior work is challenging due to fundamental 
differences in evaluation methodology. Heinrich et al.~\cite{heinrich} 
evaluate on small, input-free benchmarks using a supervised approach 
with labeled attack traces, whereas WADE operates in an unsupervised 
setting trained only on benign data. WasmEye~\cite{wasmeye}, the most 
comparable recent work, evaluates on 3 synthetic benchmarks plus 1 
real-world CVE, using a bag-of-words representation that counts function 
calls in fixed time windows without modeling their ordering. These 
fundamental differences in problem setting, data, and methodology mean 
that direct numerical comparisons would not be meaningful.

\subsection{Limitations}
\label{sec:limitations}

WADE's design entails some limitations that constrain its applicability.

First, the approach targets attacks that manifest as deviations in the 
sequence of function executions, and is therefore ineffective against 
data-only or side-channel attacks. In practice, however, such attacks 
are relatively rare against WebAssembly applications, where exploitation 
typically requires redirecting or reordering function calls to achieve 
meaningful impact.

Second, the performance of our system depends heavily on the completeness of the benign training data.  
If certain legitimate behaviors are not exercised during training, the model may incorrectly classify them as anomalies, leading to false positives.  
This issue is particularly relevant for large or input-rich applications where achieving full behavioral coverage is impractical.

Third, in real-world scenarios, the prevalence of attacks may be orders of magnitude lower than in our evaluation.
Under such extreme class imbalance, the base-rate fallacy~\cite{base-rate} 
makes the operational validity of alarms dominated by false positives: 
at a realistic attack prevalence of $0.1\%$, even a $0.42\%$ false 
positive rate yields a positive predictive value of only $\sim\!19\%$, 
meaning roughly 4 out of 5 alerts would be false alarms.

Fourth, we do not evaluate adaptive adversaries. An attacker with full
knowledge of the application may craft mimicry attacks whose function
sequences consist only of transitions observed during training, and
which would therefore produce low prediction errors and evade detection.

Despite these limitations, WADE detected all attacks in our dataset that
alter the function-level execution trace and can serve as a practical runtime protection layer for WebAssembly applications.

\section{Conclusions and Future Developments}

We presented WADE, an anomaly detection engine for browser-executed WebAssembly applications. WADE instruments WebAssembly binaries to extract function-level execution traces, encodes them as sequences, and trains a per-application LSTM model to learn normal control-flow behavior. At runtime, prediction errors exceeding adaptive thresholds trigger alerts. A multi-threshold detection mechanism with four parallel detector instances and adaptive weighted voting enables robust identification of diverse attack manifestations without prior knowledge of attack patterns.

Evaluation on 67 programs spanning ten vulnerability classes yielded zero false negatives and a 0.42\% false positive rate, confirming that sequence-based anomaly detection can effectively identify control-flow deviations in WebAssembly applications, though operational precision depends on deployment-time attack prevalence.

Two directions are planned for future work. First, extending monitoring to include memory accesses would enable detection of data-only and non-control-flow attacks that fall outside WADE's current scope. Second, expanding the dataset with real-world vulnerable WebAssembly programs would support more thorough validation across a broader range of vulnerability classes and execution contexts.

%
%
\bibliographystyle{splncs04}
\bibliography{references}

\end{document}